\documentclass{article}
\usepackage{amssymb}
\usepackage{epsfig}
\usepackage{graphicx}
\usepackage{cite}
\usepackage{amssymb}
\usepackage{rotating}
\newcommand{\bc}{\begin{center}}
\newcommand{\ec}{\end{center}}
\newcommand{\be}{\begin{equation}}
\newcommand{\ee}{\end{equation}}
\begin{document}
\bc
{\bf \Large The implications of final L3 measurement of
$\sigma_{tot}(\gamma\gamma\rightarrow b\overline{b})$}\ec
\bc{\bf\large Ji\v{r}\'{\i} Ch\'{y}la}\ec \bc{Center for Particle Physics,
Institute of Physics, Academy of Sciences of the Czech Republic\\ Na
Slovance 2, 18221 Prague 8, Czech Republic, e-mail: chyla@fzu.cz}
\ec
\bc{\large Abstract}\ec The excess of data on the total cross section of
$\overline{b}b$ production in $\gamma\gamma$ collisions over QCD
predictions, observed by L3, OPAL and DELPHI Collaborations at LEP2,
has so far defied explanation. The recent final analysis of L3 data
has brought important new information concerning the dependence of the
observed excess on the $\gamma\gamma$ collisions energy $W_{\gamma\gamma}$.
The implications of this dependence are discussed.

\vspace*{0.3cm}
In \cite{jab} we discussed various aspects of the theoretical
description of $\overline{b}b$ production in $\gamma\gamma$ collisions
which might be relevant for explanation of the excess of data on
$\sigma_{tot}(\gamma\gamma\rightarrow b\overline{b})$ over QCD
predictions, observed in \cite{L3,OPAL,DELPHI}. The conclusions
closed with the observation that in order to understand this excess
{\em ``the separation of data into at least two
bins of the hadronic energy $W_{\gamma\gamma}$, say
$W_{\gamma\gamma}\lesssim 30$ GeV and $W_{\gamma\gamma}\gtrsim 30$
GeV, could be instrumental in pinning down the possible mechanisms or
phenomena responsible for the observed excess.''}

\begin{figure}[h]\unitlength=1mm
\begin{picture}(160,60)
\put(0,0){\epsfig{file=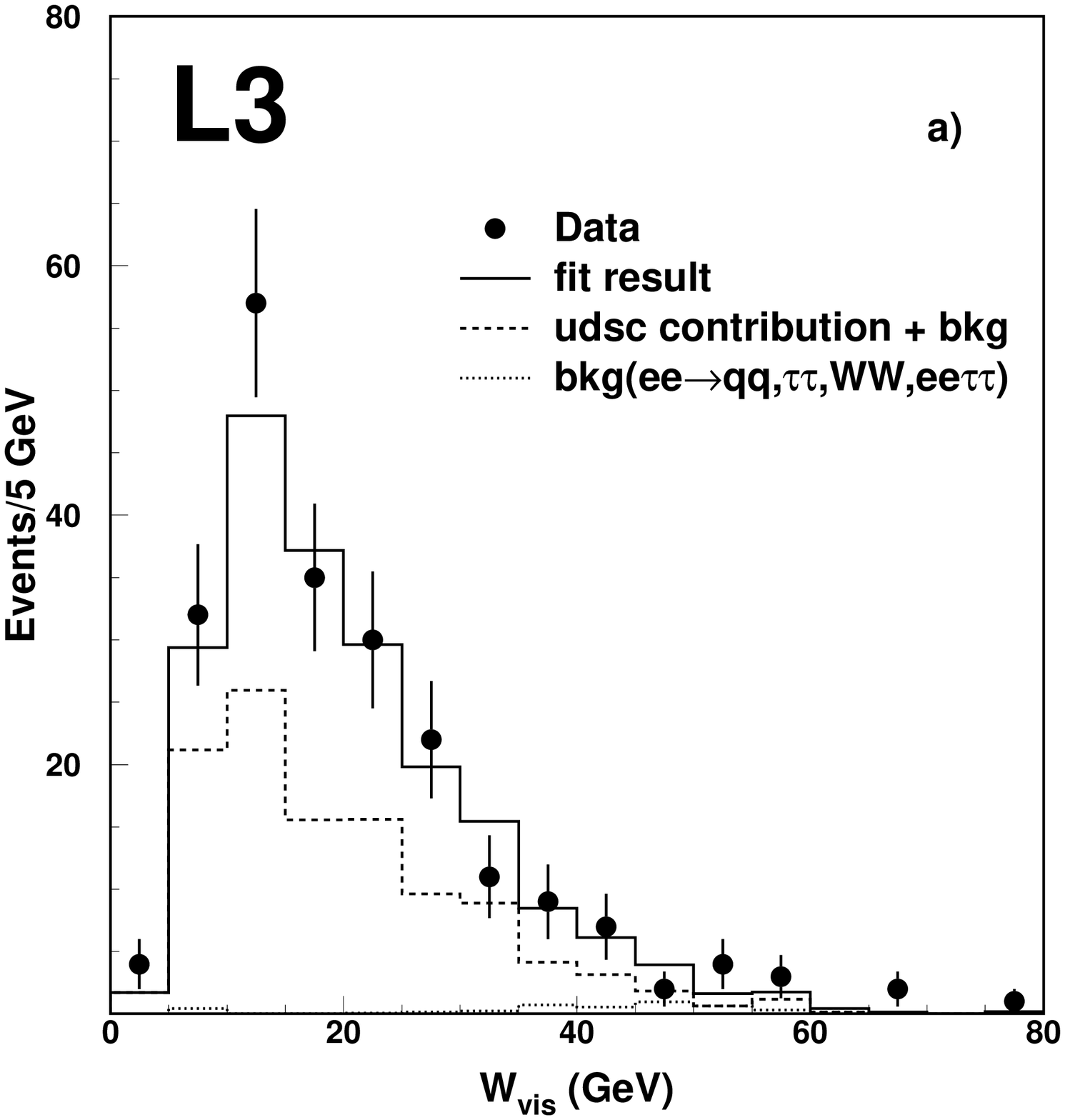,height=6.5cm}}
\put(55,0){\epsfig{file=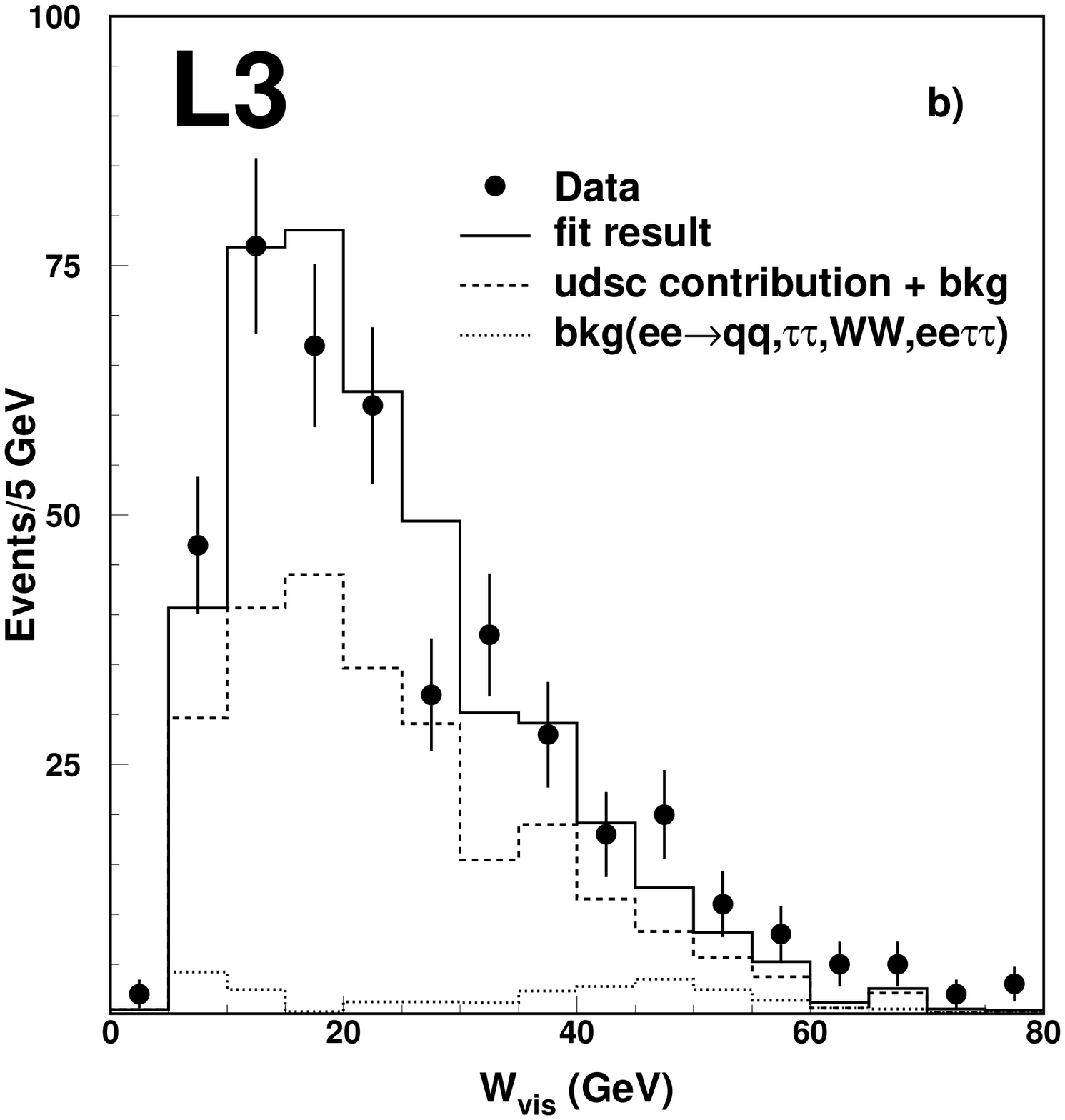,height=6.5cm}}
\put(112,0){\epsfig{file=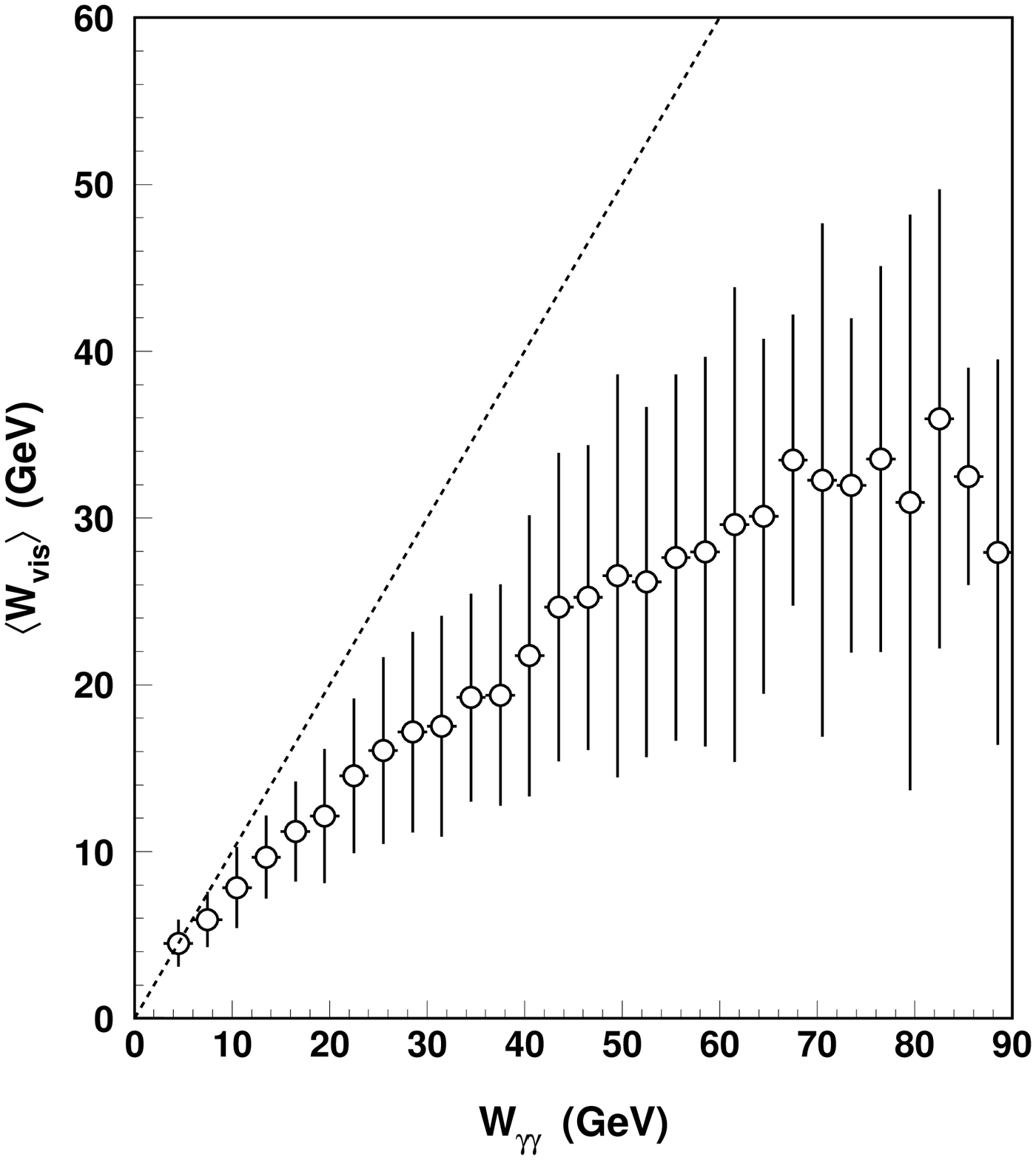,height=6.5cm}}
\end{picture}
\caption{Distribution of $\overline{b}b$ events in electron (left)
and muon (middle) samples as a function of visible energy
$W_{vis}$ observed in \cite{L3new}.
The plot on right shows the correlation between $W_{\gamma\gamma}$ and
$W_{vis}$ in $\overline{c}c$ events \cite{L3ccbar}.}
\label{Wdep}
\end{figure}

In the meantime final analysis of L3 data on the $b\overline{b}$ production
in $\gamma\gamma$ collisions at LEP2 has appeared \cite{L3new}. In this
paper the distribution of the excess of data over the theoretical
prediction \cite{zerwas} is plotted, for both the electron and muon
samples, as a function of the visible $\gamma\gamma$ collision energy
$W_{vis}$. These plots, reproduced in Fig. \ref{Wdep}a,b, show that the
excess comes predominantly from low $W_{vis}$, roughly
$W_{vis}\lesssim 30$ GeV.
To exploit this information the knowledge of the correlation between
the true $W_{\gamma\gamma}$ and the visible $W_{vis}$ energy is needed. Such
correlation was studied by L3 for the analogous case of $\overline{c}c$
production in $\gamma\gamma$ collisions in \cite{L3ccbar}. Unfortunately, no
such study is either available or foreseen \cite{Valeri} for the case
of $\overline{b}b$ production. We can therefore make only rough estimate based
on the results shown in Fig. \ref{Wdep}c. They indicate that the excess from
the region $W_{vis}\lesssim 30$ GeV translates roughly to
$W_{\gamma\gamma}\lesssim 60$ GeV, with more than half of it coming from
$W_{vis}\lesssim 20$ GeV, i.e. $W_{\gamma\gamma}\lesssim 35$ GeV. With this
information on the kinematic region wherefrom comes most of the excess measured
$\sigma_{tot}(\gamma\gamma\rightarrow b\overline{b})$ at hand, what can be
said about its possible origins?

\begin{figure}\centering
\epsfig{file=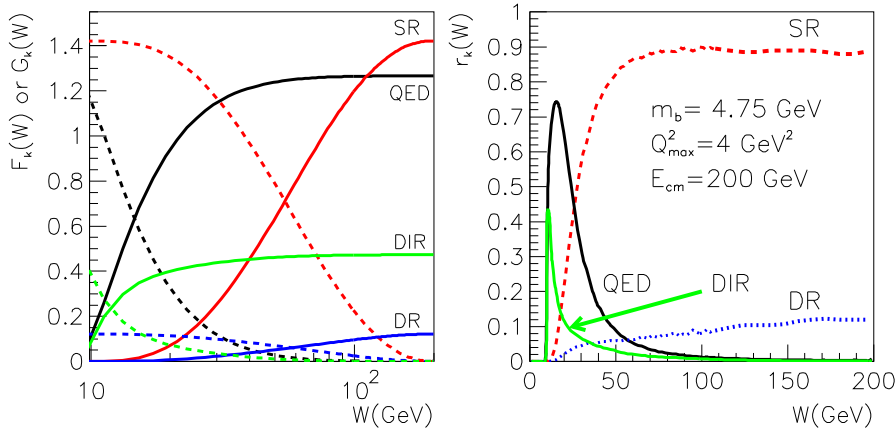,width=12cm}
\caption{Left: solid (dashed) curves show $F_k(W)$ ($G_k(W)$) for
QED and three LO QCD contributions. Right: The relative contributions
$r_k(W)$. Taken from \cite{jab}.}
\label{sgamma2}
\end{figure}

In \cite{jab} we have introduced several measures characterizing the
$W_{\gamma\gamma}$-dependence of the four individual
contributions to $\sigma_{tot}(\gamma\gamma\rightarrow b\overline{b})$:
pure QED term and three QCD corrections, direct photon and single and double
resolved photon contributions. Two of them, namely
\be
F_k(W_{\gamma\gamma})\equiv \int_{2m_b}^{W_{\gamma\gamma}}
{\mathrm{d}}w \frac{{\mathrm{d}}
\sigma_k({\mathrm{e}}^+{\mathrm{e}}^-
\rightarrow b\overline{b})}{{\mathrm{d}}w},~~
G_k(W_{\gamma\gamma})\equiv \int_{W_{\gamma\gamma}}^{\sqrt{S}}
{\mathrm{d}}w \frac{{\mathrm{d}}\sigma_k
({\mathrm{e}}^+{\mathrm{e}}^-
\rightarrow b\overline{b})
}{{\mathrm{d}}w},
\label{integ}
\ee
quantify how much of a given contribution comes from the region up to
$W_{\gamma\gamma}$ or above it, whereas $r_k(W_{\gamma\gamma})$ gives the
relative importance of individual contributions at fixed
$W_{\gamma\gamma}$. The associated plots, corresponding to pure QED term and
lowest order QCD contributions to direct and resolved photon channels
are displayed in Fig. \ref{sgamma2}. They reveal large difference in both shape
and magnitude, which, taking into account the correlation between
$W_{\gamma\gamma}$ and $W_{vis}$ can be used to draw the
following conclusions from L3 data.

\begin{figure}\centering
\epsfig{file=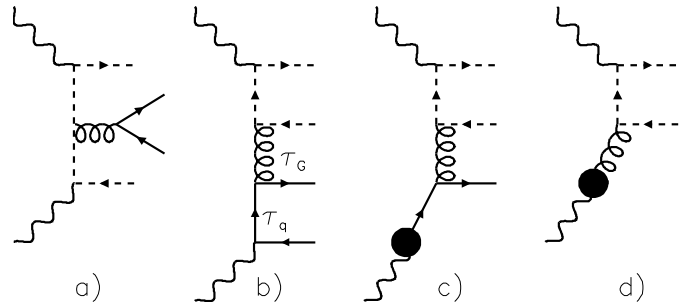,width=9cm}
\caption{a) Example of the order $\alpha^2\alpha_s^2$ diagram contributing to
direct photon component proportional to $e_b^4$. Direct photon diagram in b),
proportional to $e_b^2$, is related by factorization to single resolved photon
diagrams in c) and d). Solid (dashed) lines denote light (heavy) quarks,
filled points stand for quark and gluon distribution functions of the
resolved photon and $\tau_q$ and $\tau_G$ in b) are quark and gluon
virtualities.}
\label{examples}
\end{figure}

In the region $W_{\gamma\gamma}\lesssim 35$ GeV about 63\% of the sum of
the four contributions comes from the pure QED one, about 21\% from direct
photon and 16\% from resolved photon ones (double resolved photon contribution
is negligible throughout the kinematic region accessible at LEP2).
To enhance significantly the theoretical prediction in this region
requires very large higher QCD corrections in the latter two channels.
To my mind there are two possibilities.
As emphasized in \cite{jab} the part of $\alpha^2\alpha_s^2$ direct photon
correction proportional to $e_b^4$ (coming from diagrams like that in
Fig. \ref{examples}a), absent from all existing
calculations like \cite{zerwas}, is needed to make the direct photon
contribution of genuine next-to-leading order in $\alpha_s$. I have no idea
how large it can be but we should keep in mind that in this
range the transverse momenta of $b$ quarks are small and the hard scale
is thus given approximately by $m_b$. This is not small, but it is not large
either and so enhancement by a factor of $2-3$ does not seem impossible. Also,
the proper (in whatever sense one understands this word) choice of the
renormalization scale is likely to be crucial in this region.
Because we are close to the threshold for producing the $b\overline{b}$
pair, the threshold corrections of the type investigated in
\cite{bonciani,kidonakis} for hadroproduction of $\overline{b}b$ pairs may
also be numerically quite important. Because of a quite different initial
state in $\gamma\gamma$ collisions, it is, however, difficult to make any
quantitative guess based on such calculations.

About half of the excess comes from the region $W_{\gamma\gamma}\gtrsim 35$
GeV, where QED and direct photon contributions are negligible and the single
resolved photon dominates the sum of all lowest order contributions. As the
threshold corrections are likely to be less important here, we are
left with the question whether the existing higher order QCD corrections to
single resolved photon contribution \cite{zerwas} are reliable, i.e whether
the ``theoretical uncertainty'' attached to them correctly reflects
our (lack of) knowledge of all relevant effects.

In \cite{jab} we have shown that in this kinematic region the next-to-leading
order calculations of this contribution,
$\sigma_{\mathrm{sr}}^{\mathrm{NLO}}(W_{\gamma\gamma},M,\mu)$, considered as
a function of two independent parameters, the renormalization scale $\mu$ and
the factorization scale $M$, exhibits no region of local stability. The
standard choice of scales $\mu=M=m_b$ thus picks up a point where the
NLO results that are inherently unstable. Moreover, the standard way of
estimating the associated
``theoretical uncertainty'' by varying this common scale around $m_b$
within a factor of two is entirely ad hoc. I have furthermore argued that to
make these calculations factorization scale invariant to the order considered
the part of direct photon contribution of the order $\alpha^2\alpha_s^2$
proportional to $e_b^2$ must also be included. This contribution, which comes
from diagrams like that in Fig. \ref{examples}b integrated over the region
outside the singularities at $\tau_q=$ and $\tau_q=\tau_G=0$, is related to
the leading and next-to-leading order single resolved photon contributions
corresponding to diagrams in Fig. \ref{examples}c,d.

Unfortunately, as the mentioned direct photon calculations are not available,
we cannot check whether by adding them to the single resolved photon
contribution, which is also proportional to $e_b^2$, the sum will be more
stable than the latter contribution alone. Neither can one estimate their
numerical effect, but as in the case of the direct photon contribution of
the order $\alpha^2\alpha_s^2$ proportional to $e_b^4$, it is not
impossible that they might significantly enhance the existing NLO
calculations and bring them thus closer to the data.

In summary, the final L3 analysis of their data on
$\sigma_{tot}(\gamma\gamma\rightarrow b\overline{b})$ indicates that the
observed excess of the data over the current QCD calculations extends over
most of the accessible range, with about half of it coming from small
$W_{\gamma\gamma}\lesssim 35$ GeV. Our conjecture is that at least part
of this excess may be due to the absence of the so far uncalculated order
$\alpha^2\alpha_s^2$ direct photon contributions proportional to both
$e_b^4$ and $e_b^2$. In the region $W_{\gamma\gamma}\lesssim 35$ GeV also
the threshold corrections may be numerically important. Both of the
mentioned calculations are difficult, but certainly worth the efforts.

\vspace*{0.5cm}
\noindent
{\Large \bf Acknowledgment}

\vspace*{0.3cm}
\noindent
This work was supported by the project AV0-Z10100502 of the Academy
of Sciences of the Czech Republic and the project LC527 of the Ministry
of Education of the Czech Republic. I am grateful
to Bernard Echenard and Valeri Andreev for correspondence concerning the
correlation between true and visible $\gamma\gamma$ collision energy and
to Nikolaos Kidonakis for discussion of the importance of
threshold corrections in general hard processes.

\end{document}